\documentstyle[12pt,epsfig]{article}
\begin{document}
\begin{center}
{\Large \bf Gluon distribution at moderately low-x from NMC deuteron
structure function data}\\
\vspace{1cm}
{\large J.K. Sarma\footnote{E-mail:jks@agnigarh.tezu.ernet.in} and
G.A. Ahmed\footnote{E-mail:gazi@agnigarh.tezu.ernet.in}}\\
\vspace{.5cm}
Physics Department, Tezpur University, Napaam,\\
Tezpur-784 028, Assam, India \\
\end{center}
\vspace{1cm}

\begin{abstract}
We present some simple methods to find gluon distribution from analysis of deuteron
$F_{2}$ structure function data at moderately low-x. Here we use the leading order(LO) Altarelli
-Parisi(AP) evolution equation and New Muon Collaboration (NMC) deuteron $F_{2}$
structure function data to extract gluon distribution. We also compare our results
with those of other authors.\\ \\
\noindent \underline{PACS:}12.38.-t; 12.38.Aw; 13.60.Hb\\
\noindent \underline{Keywords:}Gluon distribution; Hadron structure; QCD; Low-x physics\\
\end{abstract}
\vspace{1cm}

\noindent {\large \bf 1.Introduction:}\\
\indent The measurements of the proton and the deuteron $F_{2}(x,Q^{2})$ structure functions by
Deep Inelastic Scattering (DIS) processes in the low-x region where $x$ is the
Bjorken variable have opened a new era in parton density measurements [1]. It is
important for understanding the inner structure of hadrons as well as examining
QCD, the underlying dynamics of strong interaction among the partons inside
hadrons. It is also important to know gluon distribution $G(x,Q^{2})$ inside
hadron at low-x because gluons are expected to be dominant in this region.
On the otherhand, measurement of gluon distribution directly from
experiments is a difficult task. It is, therefore, important to measure gluon distribution
$G(x,Q^{2})$ indirectly from proton or deuteron structure
functions $F_{2}(x,Q^{2})$. Here the representation for the gluon distribution
$G(x)=xg(x)$ is used where $g(x)$ is the gluon density. A few number of papers have already been
published [2-9] in this connection. Here we present two alternative methods to relate Gluon distribution
$G(x,Q^{2})$ with deuteron $F_{2}(x,Q^{2})$ structure function and their differential
coefficients with respect to $lnQ^{2}$ and $x$, i.e. $\partial F_{2}(x,Q^{2})/
\partial lnQ^{2}$ and $\partial F_{2}(x,Q^{2})/\partial x$ for fixed values of $Q^{2}$.
We report for the first time some methods to extract gluon distributions
from deuteron $F_{2}$ structure function data. Our methods are simpler
with less approximation and more transperent. Of course, there exist some
established methods [10] for extracting gluon distributions from data
based on global fits. In these methods, momentum distribution and other
constraints [11] are used to get gluon distributions. But our methods are
based on the direct solutions of QCD evolution equations which may be some good
alternatives.\\
\indent Section 1 in our paper is the introduction. Section 2 deals with
the theory for extracting gluon distribution from deuteron $F_{2}$
structure function data. Section 3 is the result and discussion and section 4 is the
summary and conclusion.\\
\vspace{.5cm}

\noindent{\large \bf 2.Theory:}\\
\indent In the LO analysis deuteron structure function is directly related to
singlet structure function [12]. On the otherhand, the differential coefficient
of singlet structure function $F_{2}^{s}(x,Q^{2})$ with respect to $lnQ^{2}$,
i.e. $\partial F_{2}^{s}(x,Q^{2})/\partial lnQ^{2}$ has a relation  with
singlet structure function itself as well as gluon distribution function
[12] from AP evolution equation [13-16]. Thus it is possible to calculate gluon
distribution from singlet structure function or ultimately from deuteron structure
function also.\\
\indent The LO AP evolution equation for singlet structure function [12] is
given by
\begin{eqnarray}
\lefteqn{\frac{\partial F_{2}^{s}(x,t)}{\partial t}-\frac{A_{f}}{t}[\{3+4ln(1-x)\}
F_{2}^{s}(x,t)}\nonumber\\
& & +2\int_{0}^{1-x}\frac{dz}{z}\{(z^{2}-2z+2)F_{2}^{S}\left(\frac{x}
{1-z},t\right)-2F_{2}^{s}(x,t)\}\nonumber\\
& & +\frac{3}{2}N_{f}\int_{0}^{1-x}(2z^{2}-2z+1)G\left(\frac{x}{1-z},t\right)dz]=0
\end{eqnarray}
where, $t=ln(Q^{2}/\Lambda^{2})$ and $A_{f}=4/(33-2N_{f})$, $N_{f}$ being the
number of flavours and $\Lambda$ is the QCD cut off parameter. Now,
\begin{eqnarray}
\frac{1}{1-z}=x\sum_{k=0}^{\infty}z^{k}=x+x\sum_{k=1}^{\infty}z^{k}.
\end{eqnarray}
We have, $1-x<z<0\Rightarrow|z|<1$ which implies that the expansion (2) is
convergent. Now by the Taylor expansion [17] we get, 
\begin{eqnarray}
F_{2}^{s}\left(\frac{x}{1-z},t\right)
\simeq F_{2}^{s}(x,t)+x\sum_{k=1}^{\infty}z^{k}\frac{\partial F_{2}^{s}(x,t)}
{\partial x}\\
{\rm and} \;G_{2}^{s}\left(\frac{x}{1-z},t\right)
\simeq G_{2}^{s}(x,t)+x\sum_{k=1}^{\infty}z^{k}\frac{\partial G_{2}^{s}(x,t)}
{\partial x}
\end{eqnarray}
neglecting the higher order terms.\\
\indent But as a matter of fact, we can not neglect the higher order terms
as these terms are not small in Regge-like [7,18] or in Double-logarithmical [7,19]
behaviours for singlet structure function or gluon distribution function. On
the otherhand, it has been shown that this Taylor expansion method is successfully
applied in calculating $Q^{2}$-evolution [20-22] or $x$-evolution [23] of
structure function with excellent phenomenological success. Some authors [3-5]
again applied this method to extract gluon distribution from proton structure
function. It was suggested that [23] one possible reason for success of this
method may be due to simplification of QCD processes at low-x for momentum
constraints.\\
\indent Putting equations (3) and (4) in equation (1) and performing
z-integrations we get
\begin{eqnarray}
\frac{\partial F_{2}^{s}(x,t)}{\partial t}-\frac{A_{f}}{t}[A_{s}(x)F_{2}^{s}
(x,t)+B_{s}(x)G(x,t)+C_{s}(x)\frac{\partial F_{2}^{s}(x,t)}{\partial x}\nonumber\\
+D_{s}(x)\frac{\partial G(x,t)}{\partial x}]=0,\\
{\rm where,}\; A_{s}(x)=3+4ln(1-x)+2\{(1-x)(-2+(1-x)/2)\},\nonumber\\
B_{s}(x)=(3/2)N_{f}\{(1-x)(x+(2/3)(1-x)^{2})\},\nonumber\\
C_{s}(x)=2x\{ln(1/x)+(1-x)(1-(1-x)/2)\}\nonumber\\
{\rm and}\; D_{s}(x)=(3/2)N_{f}\{ln(1/x)-(1-x)(1+(2/3)(1-x)^{2})\}.\nonumber
\end{eqnarray}
Now, we can apply two methods to extract gluon distributions:\\ \\
\noindent \underline{\bf First method:}\\
\indent At very low-x limit, $x\rightarrow 0$, the functions $A_{s}(x)$,
$C_{s}(x)$, and $D_{s}(x)$ become vanished and $B_{s}(x)=N_{f}$. Equation (5) then
becomes simplified and we get
\begin{eqnarray}
\frac{\partial F_{2}^{s}(x,t)}{\partial t}-\frac{A_{f}}{t}.N_{f}.G(x,t)=0\nonumber\\
\Rightarrow G(x,t)=\frac{t}{A_{f}N_{f}}.\frac{\partial F_{2}^{s}(x,t)}{\partial t}.
\end{eqnarray}
Equation (6) is a very simple relation between gluon distribution function with the
differential coefficient of singlet structure function with respect to $t$.\\ \\
\noindent \underline{\bf Second method:}\\
\indent Recasting equation (5) we get
\begin{eqnarray}
G(x,t)+\frac{D_{s}(x)}{B_{s}(x)}\frac{\partial G(x,t)}{\partial x}=\frac{1}
{A_{f}B_{s}(x)}.t.\frac{\partial F_{2}^{s}(x,t)}{\partial t}\nonumber\\
-\frac{A_{s}(x)}{B_{s}(x)}F_{2}^{s}(x,t)-\frac{C_{s}(x)}{B_{s}(x)}\frac{\partial F_{2}^{s}(x,t)}
{\partial x}.
\end{eqnarray}
Now $D_{s}(x)/B_{s}(x)$ is very small at low-x, $lim_{x\rightarrow 0}D_{s}(x)/B_{s}(x)=0.$
So, applying the Taylor expansion series we can write
\begin{eqnarray}
G(x,t)+\frac{D_{s}(x)}{B_{s}(x)}\frac{\partial G(x,t)}{\partial x}
=G\left(x+\frac{D_{s}(x)}{B_{s}(x)},t\right).\nonumber
\end{eqnarray}
\indent Thus equation (7) gives
\begin{eqnarray}
G(x',t)=K_{1}(x).t.\frac{\partial F_{2}^{s}(x,t)}{\partial t}
+K_{2}(x)\frac{\partial F_{2}^{s}(x,t)}{\partial x}+K_{3}(x)F_{2}^{s}(x,t),\\
{\rm where,}\;x'=x+\frac{D_{s}(x)}{B_{s}(x)},\; K_{1}(x)=\frac{1}{A_{f}B_{s}},\nonumber\\
K_{2}(x)=-\frac{C_{s}(x)}{B_{s}(x)} \;{\rm and}\; K_{3}(x)=-\frac{A_{s}(x)}{B_{s}(x)}.\nonumber
\end{eqnarray}
Equation (8) is also a simple relation between gluon distribution function
with the differential coefficients of singlet structure function with respect to
$t$ and $x$, and with singlet structure function itself. If we try to combine the
last two terms of equation (8) let us take common $K_{3}(x)$ from both the
terms which reduce to
\begin{eqnarray}
K_{3}(x)\left[F_{2}^{s}(x,t)+\frac{K_{2}(x)}{K_{3}(x)}.\frac{\partial F_{2}^{s}(x,t)}{\partial x}\right].\nonumber
\end{eqnarray}
But $K_{2}(x)/K_{3}(x)$ is not small at low-x and therefore these two terms
can not be combined to one as in the case of gluon by applying Taylor expansion series.\\
\indent The relation between deuteron and singlet structure function at
LO [12] is
\begin{eqnarray}
F_{2}^{d}(x,t)=\frac{5}{9}F_{2}^{s}(x,t)\Rightarrow F_{2}^{s}(x,t)
=\frac{9}{5}F_{2}^{d}(x,t).
\end{eqnarray}
Then we get
\begin{eqnarray}
\frac{\partial F_{2}^{s}(x,t)}{\partial t}=\frac{9}{5}.\frac{\partial F_{2}^{d}(x,t)}{\partial t}\\
{\rm and}\;\frac{\partial F_{2}^{s}(x,t)}{\partial x}=\frac{9}{5}.\frac{\partial F_{2}^{d}(x,t)}{\partial x}.
\end{eqnarray}
Putting equations (9), (10) and (11) in equations (6) and (8), we get repectively
\begin{eqnarray}
G(x,t)=\frac{9t}{5A_{f}N_{f}}.\frac{\partial F_{2}^{d}(x,t)}{\partial t}\;\rm{and}\\
{G(x',t)}
=\frac{9}{5}\left[K_{1}(x).t.\frac{\partial F_{2}^{d}(x,t)}{\partial t}
+K_{2}(x)\frac{\partial F_{2}^{d}(x,t)}{\partial x}+K_{3}(x)F_{2}^{d}(x,t)\right],
\end{eqnarray}
which are our main results. From these equations it is seen
that if we have deuteron structure function and their differential coefficients
with respect to $t$ and $x$ at any $x$ for a fixed value of $t=t_{0}$, we
can calculate gluon distribution function at $x$ (first method) from
equation (12) or at $x'=x+D_{s}(x)/B_{s}(x)$ (second method) from equation (13)
as a LO analysis.\\
\indent For analysis of our results, we use NMC 15-parameter function [24,25]
which parametrized their data for proton and deuteron structure functions
for $Q^{2}$ values from $0.5\;GeV^{2}$ to $75\;GeV^{2}$ and low-x values from 0.002
to 0.6. This parametrization can also well describe the SLAC [26] and
BCDMS [27] data, and Fermilab [28] low-x data. The function
used to describe proton as well as deuteron data is given by
\begin{eqnarray}
F_{2}(x,Q^{2})=A(x)\left[\frac{ln(Q^{2}/\Lambda^{2})}{ln(Q_{0}^{2}/\Lambda^{2})}\right]^{B(x)}.
\left[1+\frac{C(x)}{Q^{2}}\right].
\end{eqnarray}
Here $Q_{0}^{2}=20\;GeV^{2}$ and $\Lambda=0.250\;GeV$,\\
$A(x)=x^{a_{1}}(1-x)^{a_{2}}\{a_{3}+a_{4}(1-x)+a_{5}(1-x)^{2}+a_{6}(1-x)^{3}
+a_{7}(1-x)^{4}\}$,\\
$B(x)=b_{1}+b_{2}x+b_{3}/(x+b_{4})$ and\\
$C(x)=c_{1}x+c_{2}x^{2}+c_{3}x^{3}+c_{4}x^{4}$\\
where $a_{1}$, $a_{2}$, $a_{3}$, $a_{4}$, $a_{5}$, $a_{6}$, $a_{7}$, $b_{1}$,
$b_{2}$, $b_{3}$, $b_{4}$, $c_{1}$, $c_{2}$, $c_{3}$ and $c_{4}$ are the 15-parameters
used to fit the data. Actually two different sets of these parameters are used
to describe proton and deuteron structure functions in the same equation,
equation (14). Thus for the respective sets of parameters equation (14) gives
the deuteron structure function as
\begin{eqnarray}
F_{2}^{d}(x,t)=A(x)\left[\frac{t}{t_{0}}\right]^{B(x)}.\left[1+\frac{e^{-t}}{\Lambda^{2}}.C(x)\right].
\end{eqnarray}
where $t=ln(Q^{2}/\Lambda^{2})$ and $t_{0}=ln(Q_{0}^{2}/\Lambda^{2})$.
Differentiating $F_{2}^{d}(x,t)$ with respect to $t$ and $x$ we get respectively
\begin{eqnarray}
\frac{\partial F_{2}^{d}(x,t)}{\partial t}=\left(\frac{B(x)}{t}-1\right)F_{2}^{d}(x,t)
+A(x)\left(\frac{t}{t_{0}}\right)^{B(x)}\;\rm{and}\\
\frac{\partial F_{2}^{d}(x,t)}{\partial x}
=\left[\left(\frac{t}{t_{0}}\right).\frac{\partial A(x)}{\partial x}
+A(x)\left(\frac{t}{t_{0}}\right)^{B(x)}
.ln\left(\frac{t}{t_{0}}\right).\frac{\partial B(x)}{\partial x}\right].\nonumber\\
\left[1+\frac{e^{-t}}{\Lambda^{2}}.C(x)\right]
+A(x)\left(\frac{t}{t_{0}}\right)^{B(x)}
.\left[\frac{e^{-t}}{\Lambda^{2}}.\frac{\partial C(x)}{\partial x}\right]
\end{eqnarray}
where,
$$\frac{\partial A(x)}{\partial x}=\left(\frac{a_{1}}{x}-\frac{a_{2}}{1-x}\right)A(x)
-x^{a_{1}}(1-x)^{a_{2}}.\{a_{4}+2a_{5}(1-x)$$
$$+3a_{6}(1-x)^{2}+4a_{7}(1-x)^{3}\},$$
$$\frac{\partial B(x)}{\partial x}=b_{2}-\frac{b_{3}}{(x+b_{4})^{2}}\;\rm{and}$$
$$\frac{\partial C(x)}{\partial x}=c_{1}+2c_{2}x+3c_{3}x^{2}+4c_{4}x^{3}.$$
Now putting equations (15), (16) and (17) in equatins (12) and (13) we can
easily calculate gluon disatributions at $x$ (first method) or
$x'=x+D_{s}(x)/B_{s}(x)$ (second method) respectively.\\
\vspace{.5cm}

\noindent{\large \bf 3.Results and Discussion:}\\
\indent The NMC 15-parameter function [24,25] we use, parametrized the NMC
for $Q^{2}$ values from $0.5\;GeV^{2}$ to $75\;GeV^{2}$ and low-x values from 0.002
to 0.6 which also well describe the SLAC [26] and
BCDMS [27] data and Fermilab [28] low-x data. As the data range of x we use is
moderatelt low, we will restrict our analysis for $Q^{2}$ values from $10\;GeV^{2}$ to
$60\;GeV^{2}$ and low-x values from .1 to 0.001. We can not extend our analysis
to HERA low-x region [1] due to lack of deuteron $F_{2}$ structure function
data in that region.\\
\indent In Fig.1(a) and Fig.1(b) gluon distributions obtained by our first
method (equation (12)) from NMC deuteron parametrization from the 15-parameter
function are presented at $Q^{2}=10\;GeV^{2}$ and $60\;GeV^{2}$ respectively.
The middle lines are the results without  considering the error. The upper and
the lower lines are the results with parameter values by adding and subtracting
the statistical and the systematic errors with the middle values respectively.
It has been seen that the middle lines almost coincide with the upper ones.
We calculate gluon distributions for x-values from $10^{-1}$ to $10^{-3}$
for both $Q^{2}=10\;GeV^{2}$ and $Q^{2}=60\;GeV^{2}$. In both the cases
$G(x,Q^{2})$ values increases when $x$ decreases as expected, but $G(x,Q^{2})$
is higher in $Q^{2}=60\;GeV^{2}$ than in $Q^{2}=10\;GeV^{2}$ for same $x$
especially in lower-x side. Moreover, rate of increment of $G(x,Q^{2})$ is
very high from $x=10^{-1}$ to $10^{-2}$. But the rate decreases to some
extent to lower-x region.\\
\indent Exactly in the similar way, in Fig.2(a) and Fig.2(b) gluon distributions obtained by our
second method (equation (13)) from NMC deuteron parame-trization from the 15-parameter
function are presented at $Q^{2}=10\;GeV^{2}$ and $60\;GeV^{2}$ respectively.
All discussion are exactly same as for Fig.1(a) and Fig.1(b). But overall values of
$G(x,Q^{2})$ are higher in second method than in first one for any value of $x$.
For example, $G(x,Q^{2})$ medium values are almost 20\% and 25\% higher in
second method than in first method for $Q^{2}=10\;GeV^{2}$ and
$Q^{2}=60\;GeV^{2}$ respectively at $x=10^{-3}$. This is because in our first method
we apply very low-x approximation and neglected $A_{s}(x)$, $C_{s}(x)$ and $D_{s}(x)$
in equation (5) as they are vanishingly small at very low-x to obtain equation
(6) and then equation (12). On the otherhand, in our second method we
do not apply such approximation and automatically the contributions from these
functions have been included in equation (13).\\
\indent In Fig.3, comparison of gluon distributions obtained by Sarma and Medhi method (SM),
Bora and Choudhury method (BC), Prytz method, our first method (SA 1st) and
our second method (SA 2nd) is presented for
middle values only for $Q^{2}=60\;GeV^{2}$. Values are higher for the results
of other authors with proton structure function
data than of ours with deuteron structure functions data.
This is actually due to the fact that the scaling violations of deuteron
structure functions $F_{2}^{d}(x,Q^{2})$ with respect to $lnQ^{2}(\equiv t)$
are themselves considerably less than those of HERA proton data due to H1 [29,31]
and ZEUS [30,32]
collaborations and these scaling violations are directly proportional to gluon distributions
in the formulas used by BC and Prytz to calculate gluon distributions. These HERA proton data covers
x values upto at least $\sim 10^{-4}$ in comparison with those of NMC data
which covers upto $\sim 10^{-3}$ only.\\
\indent Gluon distributions increase as $x$ decreases due to
all the authors as expected from QCD analysis. Moreover,
gluon distribution by our first method is lowest and by Sarma and Medhi method is the
highest for a particular low-x.\\
\vspace{.5cm}

\noindent{\large \bf 4.Summary and Conclusion:}\\
\indent In this article we present for the first time a method to extract
gluon distribution from the measurement of moderately low-x deuteron structure functions
and their differential coefficients with respect to $t\equiv lnQ^{2}$ and
$x$. Here we use LO AP evolution equation to relate gluon distribution function
with moderately low-x structure functions or their differential coefficients.
In our analysis we  use only NMC deuteron data parametrization by a 15-parameter
function. We find that gluon distribution from deuteron also increases when
$x$ decreases as in the case of proton as usual. We can not compare our result
of NMC data with other because low-x deuteron data is not sufficiently available.
Moreover, no other work to calculate gluon distribution function from deuteron
data has been so far reported. But we compare our result with gluon distributions
due to  other authors Sarma and Medhi, Bora and Choudhury and Prytz calculated from
low-x proton data. We see that our result is to some extent less
as differential coefficient of deuteron structure function with respectr to
$t\equiv lnQ^{2}$ is much less than that of proton structure function.\\
\indent In our method the first order approximation in Taylor expansion in
$F_{2}^{s}(x/(1-z),t)$ and  $G(x/(1-z),t)$ is used, i.e. only terms having first order
differential coefficients  $\partial F_{2}^{s}(x,t)/\partial x$ and
$\partial G(x,t)/\partial x$ are used. Scope is still there to include higher
order terms of the Taylor expansion series. Of course, our preliminary work
including second order differential coefficients $\partial^{2}F_{2}^{s}(x,t)/\partial x^{2}$
and $\partial^{2}G(x,t)/\partial x^{2}$ reveals that they could not contribute
significantly. Moreover, this is only a LO analysis. We can expect better
result if we include next-to-leading order (NLO) and subsequent terms
in perturbative QCD. Work is going on in this regards.\\
\vspace{.5cm}

\noindent {\large \bf References:}\\
$[1]$See for example, W. Buchmuller and G. Inglelman, eds., Proc. Workshop
`Physics at HERA', Hamburg(1991).\\
$[2]$A.M. Cooper-Sarkar et. al., Z. Phys. C39(1988)281.\\
$[3]$K. Prytz, Phys. Lett. B311(1993)286.\\
$[4]$K. Prytz, Phys Lett. B332(1994)393.\\
$[5]$Kalpana Bora and D.K. Choudhury, Phys. Lett. B354(1995)151.\\
$[6]$R.K. Ellis, Z. Kunszt and E.M. Levin, Nucl. Phys. B420(1994)517.\\
$[7]$A.V. Kotikov, G. Parente, Phys. Lett. B379(1996)195.\\
$[8]$A.V. Kotikov, Phys. Rev. D49(1994)5746.\\
$[9]$J.K. Sarma and G.K. Medhi, Euro. Phys. J. C(in press).\\
$[10]$A.D. Martin et. al., hep-ph/9803445 (1998).\\
$[11]$A.D. Martin and F. Halzen, `Quarks and Leptons', John Wiley and Sons.,
New York(1990).\\
$[12]$L.F. Abbott, W.B. Atwood and R.M. Barnett, Phys. Rev. D22(1980)582.\\
$[13]$G. Altarelli and G. Parisi, Nucl. Phys. B12(1997)298. \\
$[14]$G. Altarelli, Phys. Rep. 81(1981)1.\\
$[15]$V.N. Gribov and L.N. Lipatov, Sov. J.Nucl. Phys. 209(1975)94.\\
$[16]$Y.L. Dokshitzer, Sov. Phys. JETP 46(1977)641.\\
$[17]$I.S. Grandshteyn and I.M. Ryzhik, `Tables of Integrals, Series and
Products' ed., A. Jeffrey, Academic Press, New York, (1965).\\
$[18]$P.D.B. Collins, `An Introduction to Regge Theory and High-Energy Physics',
Cambridge University Press, Cambridge (1977).\\
$[19]$R.D. Ball and S. Forte, Phys. Lett. B335(1994)77; B336(1994)77. \\
$[20]$D.K. Choudhury and A. Saikia, Pramana--J.Physics 29(1987)385;\\
33(1989)359; 34(1990)85.\\
$[21]$D.K. Choudhury and J.K Sarma, Pramana--J.Physics. 38(1992)481; 39(1992)273.\\
$[22]$J.K. Sarma and B. Das, Phy. Lett. B304(1993)323.\\
$[23]$J.K. Sarma, D.K. Choudhury, G.K. Medhi, Phys. Lett. B403(1997)139.\\
$[24]$M. Arneodo et. al., NMC, Phys. Lett. B364(1995)107. \\
$[25]$M. Arneodo et. al., NMC, Nucl. Phys. B483(1997)3. \\
$[26]$L.W. Whitlow et. al., Phys. Lett. B282(1992)475. \\
$[27]$A.C. Benvenuti et. al., BCDMS Collaboration, Phys. Lett.\\
B233(1989)485; 237(1990)592.\\
$[28]$A.V. Kotwal, Ph.D. thesis, Harvard University (1995).\\
$[29]$S. Aid et. al., H1 collaboration, Nucl.Phys. B470(1996)3.\\
$[30]$M. Derrick et. al., ZEUS collaboration, DESY 96-076 (1996).\\
$[31]$S. Aid et. al., H1 collaboration, Phys. Lett. B354(1995)494.\\
$[32]$M. Derrick et al, ZEUS collaboration, Phys. Lett. B364(1995)576. \\

\newpage

\begin{figure}[h]
    \begin{picture}(400,180)(0,0)
        \put (40,0){\includegraphics*[0mm,0mm][121mm,75mm]{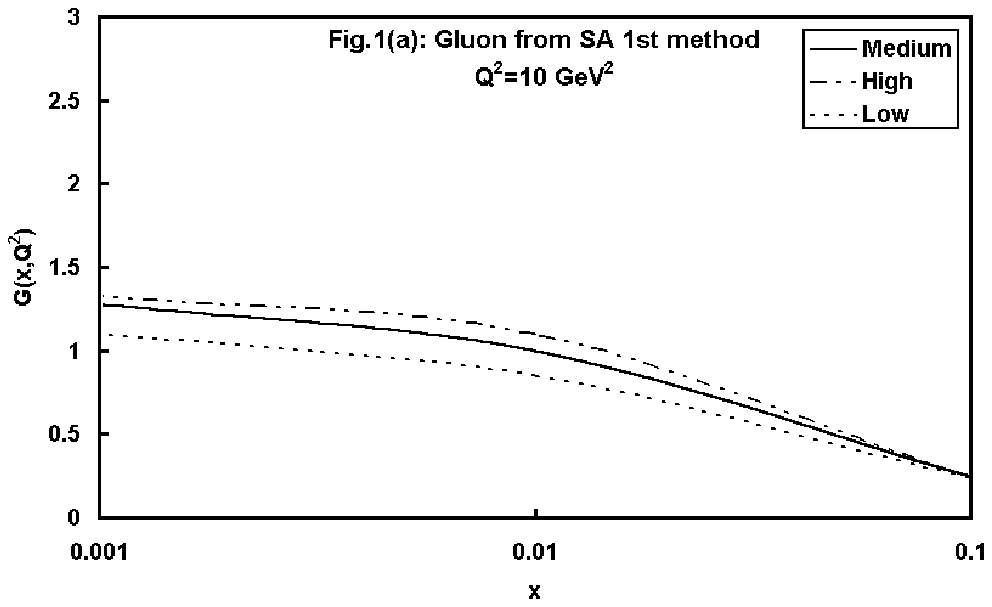}}
    \end{picture}
\end{figure}

\begin{figure}[h]
    \begin{picture}(400,180)(0,0)
        \put (40,0){\includegraphics*[0mm,0mm][121mm,75mm]{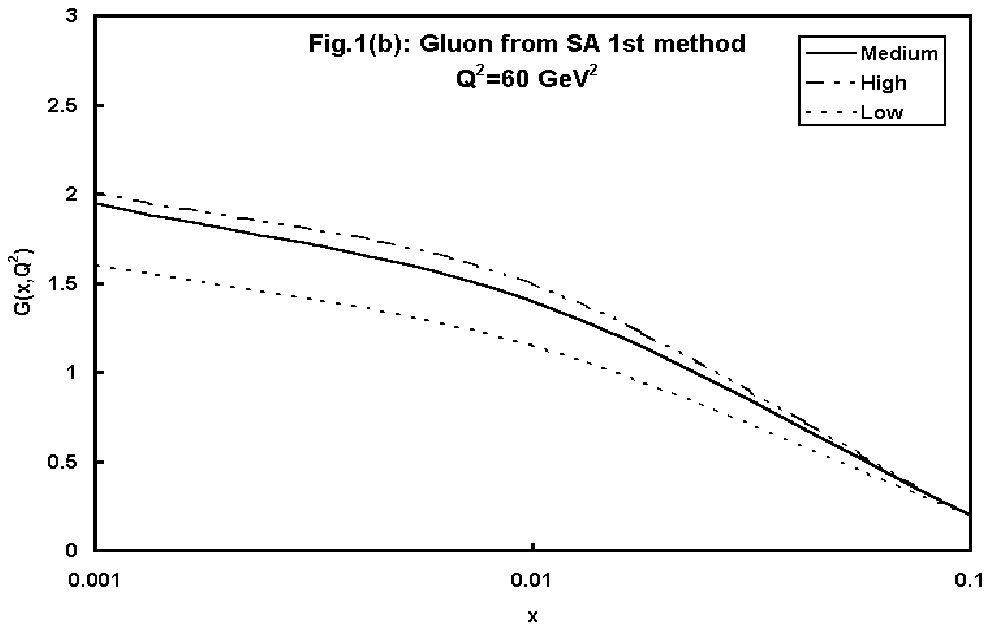}}
    \end{picture}
\end{figure}

\noindent {\underline{\bf Fig.1(a) and Fig.1(b):}}Gluon distributions obtained by our first
method (equation (12)) from NMC deuteron parametrization from the 15-parameter
function are presented at $Q^{2}=10\;GeV^{2}$ and $60\;GeV^{2}$ respectively.
The middle lines are the results without  considering the error. The upper and
the lower lines are the results with parameter values by adding and subtracting
the statistical and the systematic errors with the middle values respectively.\\ 

\newpage

\begin{figure}[h]
    \begin{picture}(400,180)(0,0)
        \put (40,0){\includegraphics*[0mm,0mm][121mm,75mm]{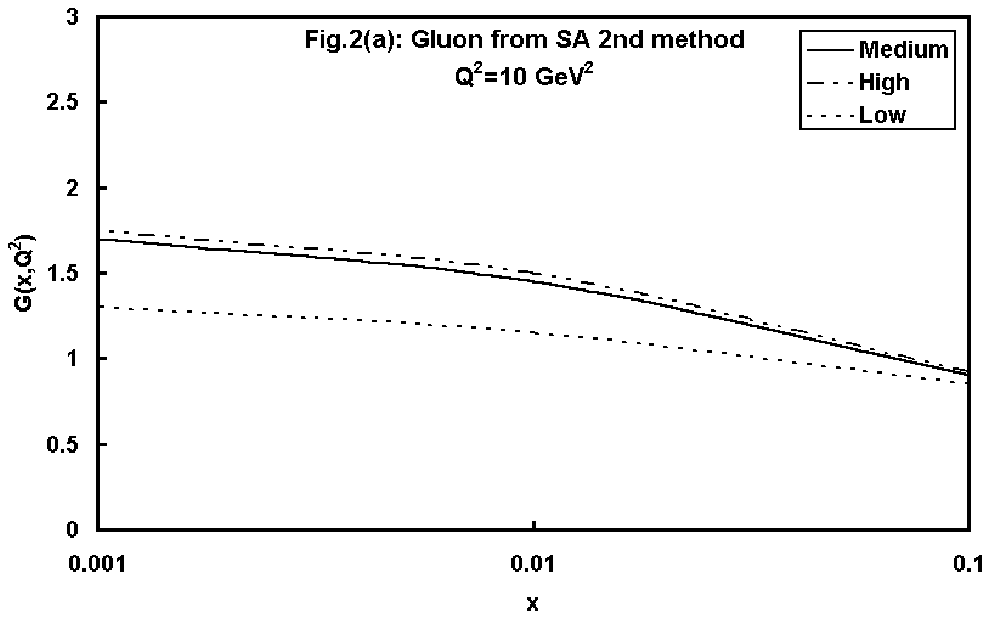}}
    \end{picture}
\end{figure}

\begin{figure}[h]
    \begin{picture}(400,180)(0,0)
        \put (40,0){\includegraphics*[0mm,0mm][121mm,75mm]{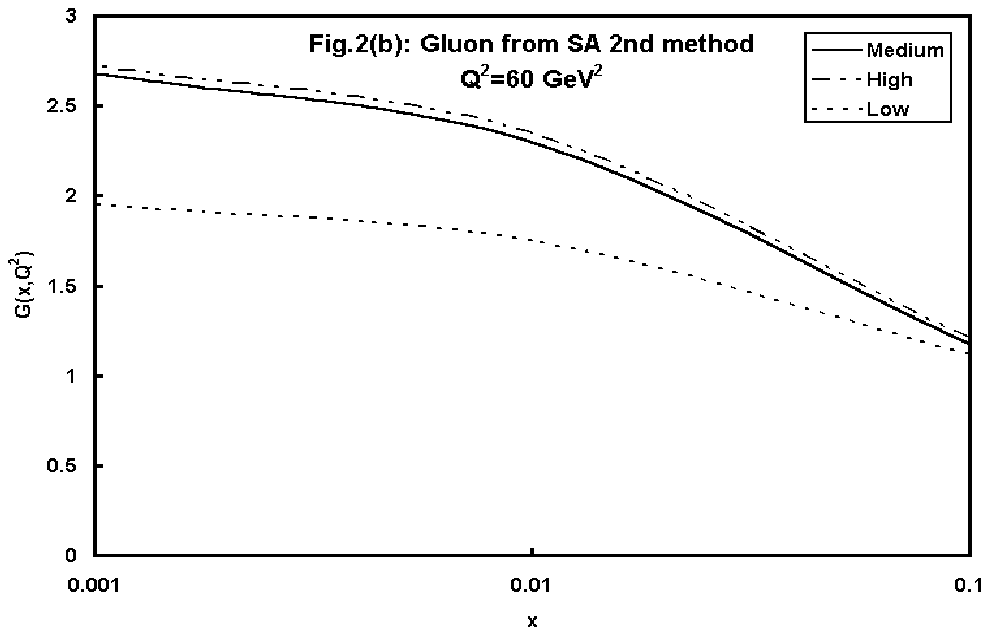}}
    \end{picture}
\end{figure}
\noindent {\underline{\bf Fig.2(a) and Fig.2(b):}}Same result as in Fig.1(a) and Fig.1(b)
by our second method (equation (13)).\\

\newpage

\begin{figure}[h]
    \begin{picture}(400,180)(0,0)
        \put (40,0){\includegraphics*[0mm,0mm][121mm,75mm]{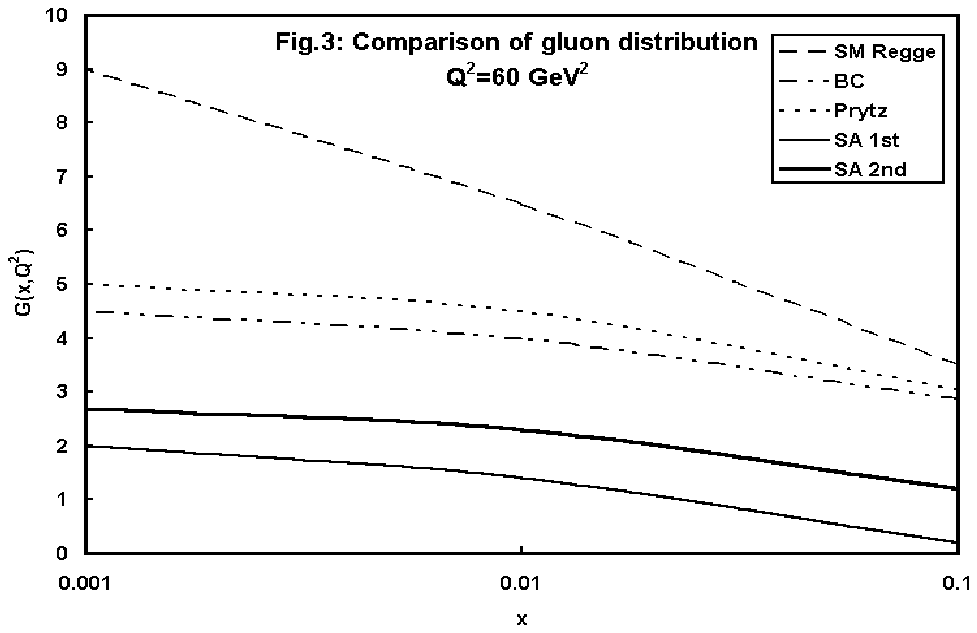}}
    \end{picture}
\end{figure}

\noindent {\underline{\bf Fig.3:}}Comparison of gluon distributions obtained by Sarma and Medhi method (SM),
Bora and Choudhury method (BC), Prytz method, our first method (SA 1st) and
our second method (SA 2nd) for middle values only
without considering any error.
\end{document}